\documentclass[12pt,reqno,a4paper]{amsart}
\newtheorem{thm}{Theorem}[section]

\newtheorem{cor}{Corollary}[section]
\newtheorem{pr}{Proposition}[section]
\theoremstyle{definition}
\newtheorem{rem}{Remark}[section]

\newcommand{\be}{\begin{equation}}
\newcommand{\ee}{\end{equation}}
\newcommand{\bea}{\begin{eqnarray}}
\newcommand{\eea}{\end{eqnarray}}
\newcommand{\beb}{\begin{eqnarray*}}
\newcommand{\eeb}{\end{eqnarray*}}
\usepackage{amssymb,amsmath, amsfonts,amsthm,setspace,indentfirst,mathrsfs}
\usepackage{xcolor,url}
\usepackage{minibox}

\usepackage{enumitem}
\usepackage{cite}
\numberwithin{equation}{section}

\begin{document}

\title{Sultana--Dyer Black Hole Spacetime and $h$-Almost Yamabe Solitons}
\author[A. A. Shaikh and Kamiruzzaman]{Absos Ali Shaikh$^{*1}$ and Kamiruzzaman$^2$}
\date{\today}

\address{\noindent\newline$^{1,2}$ Department of Mathematics,
	\newline University of Burdwan, 
	\newline Golapbag, Burdwan-713104,
	\newline West Bengal, India} 
\email{aashaikh@math.buruniv.ac.in$^1$, aask2003@yahoo.co.in$^1$}
\email{kamiruzzaman8145@gmail.com$^2$}
\begin{abstract}
This paper investigates Yamabe-type soliton structures on the Sultana–Dyer (SD) black-hole spacetime. By employing the explicit metric and curvature quantities associated with the SD spacetime, we demonstrate that the non-Killing vector field $\partial_t$ induces an $h$-almost Yamabe soliton for explicitly determined functions $h$ and $\lambda$. We also establish the condition under which this structure reduces to an almost Yamabe soliton and examine why the coordinate vector fields $\partial_r$ and $\partial_\theta$ do not satisfy the corresponding Yamabe flow conditions.
Furthermore, we show that the energy-momentum tensor, consisting of a timelike dust component and a null radiation fluid, is compatible with the Riemann, conformal (Weyl), concircular, conharmonic, and projective curvature structures. This compatibility reveals a strong geometric correspondence between the matter distribution and the causal structure of the spacetime. However, when the time-dependent scalar-curvature profile is subject to the constraints associated with late-stage cosmic expansion, the null, weak, and dominant energy conditions are violated. These findings provide a geometric characterization of the time-dependent Sultana–Dyer spacetime within the framework of generalized Yamabe solitons.
\end{abstract}

\noindent\footnotetext{$^*$Corresponding author(Absos Ali Shaikh, E-mail address: aashaikh@math.buruniv.ac.in, aask2003@yahoo.co.in).\\
$\mathbf{2020}$\hspace{5pt}Mathematics\; Subject\; Classification: 53B20; 53B30; 53B50; 53C15; 53C24; 53C65.}
\maketitle

\section{Introduction and preliminaries}\label{intro}

Geometric structures on spacetimes provide a useful framework for studying the interaction between gravitational geometry and cosmological evolution. The Sultana--Dyer (SD) spacetime is of particular interest because it describes a time-dependent cosmological black hole obtained from the Schwarzschild geometry by a conformal transformation. It therefore provides a natural setting in which black hole geometry and cosmological expansion can be studied simultaneously.\\
\indent The curvature of spacetime is a direct manifestation of gravity in general relativity. Determining the structural history of the cosmos and its underlying mathematical foundation requires analyzing curvature qualities across several spacetime models. Curvature analysis, where metrics like Schwarzschild and Kerr characterize extreme phenomena like event horizons and frame-dragging, is crucial to understanding gravitational dynamics. Curvature tensors monitor the spread of gravitational radiation from enormous cosmic collisions, whereas the Friedmann-Lema\^{i}tre-Robertson-Walker (FLRW) metric represents universal expansion on a cosmic scale based on global curvature parameters.

Beyond observable phenomena, curvature properties are foundational for probing theoretical frameworks. Specific curvature conditions enabled the formulation of singularity theorems, demonstrating the inevitability of singularities during generic gravitational collapse. Comparative curvature analysis rigorously tests general relativity against alternative paradigms like $f(R)$ gravity, and investigating micro-scale dynamics remains crucial for formulating quantum gravity. Simultaneously, these physical models promote pure mathematics by offering practical applications that propel advancements in pseudo-Riemannian geometry, employing curvature invariants for the coordinate-independent categorization of various cosmic topologies.\\
\indent The sophisticated geometric structures of certain black hole spacetimes, such as the Bardeen, Hayward, Lemos, and interior black hole metrics, are profoundly revealed by the mathematical studies conducted by Shaikh and collaborators. Through a thorough analysis of the interactions and non-commutation characteristics of various curvature tensors, including the Riemann, Ricci, Weyl, and conharmonic tensors, this study outlines the exact processes through which structural curvature regularizes. For normal black holes, which theoretically avoid core infinite singularities, this geometric classification is very important. Complex spacetime dynamics, such as the temporal evolution of mathematical symmetries within the interior metric where spatial and temporal coordinates invert over the event horizon, are also clarified by examining these curvature features.

These studies demonstrate that such black hole metrics show several classes of pseudosymmetry beyond structural regularity, resulting in a coordinate-independent physical profile. This shows that under certain geometric transformations, the curvature tensors exhibit precise proportionalities even in the absence of perfect symmetry. Validating the viability of these metrics within modified gravity paradigms requires proving that they satisfy sophisticated algebraic criteria, such as operating as a generalized Roter-type space or an Einstein manifold of level 2. Ultimately, by allowing for atypical matter fields and higher-curvature dynamics, this rigorous geometric foundation connects classical general relativity with more extensive theoretical frameworks, such as string theory and quantum loop gravity.\\
\indent The relationship between the Ricci tensor, the energy-momentum tensor (EMT), and other fundamental curvature tensors in the context of general relativity offers significant insights into the physical structure of spacetime. The alignment of these tensors reveals a smooth coherence when examining the Sultana-Dyer spacetime, which represents a dynamic Schwarzschild black hole immersed in an expanding cosmological background. In particular, the causal and gravitational structure of the underlying spacetime is fully synchronized with the matter fields causing cosmic expansion.

These geometric formations' mathematical compatibility translates into extremely particular physical properties. Algebraic purity inside the curvature is ensured by alignment with the Riemann tensor, which eliminates topological flaws and provides a stable gravitational environment. Additionally, null geodesics and angles are preserved by conformal (or Weyl) compatibility, which forces the spacetime into an algebraically special classification where the magnetic component vanishes. As a result, there is no frame-dragging radiation and only a gravito-electric field. Stellar orbits can follow smoothly because the concircular tensor connects geodesic circles, bridging the uniform expansion of the universe with the strong local gravity of the black hole. Furthermore, regardless of the temporal parameterization used, compatibility with the projective tensor ensures that unparameterized geodesic routes are established, which guarantees that the physical arrangement of matter naturally conforms to the trajectories of light and free-falling bodies. Lastly, zero-rest-mass scalar fields can propagate cleanly without scattering distortions because the conharmonic tensor preserves the integrity of harmonic functions.

A homogeneous perfect fluid (which represents cosmic dust) and an inhomogeneous directed radiation flow (a null fluid) make up the dynamically complicated matter distribution found in the Sultana-Dyer model. Several significant physical insights result from this multi-component EMT's simultaneous harmony with all five primary curvature tensors. First, it shows a strict symmetry alignment: the universal expansion and the black hole's confined gravitational pull coexist together without causing geometric shear or structural stress. It also guaranties causal consistency. The underlying conformal transformations maintain a stable causal structure and event horizon that closely resemble a typical, well-behaved Schwarzschild black hole despite the dynamic cosmic environment. Lastly, the lack of Weyl magnetic components validates the metric as an accurate and realistic cosmological model by confirming that the cosmic fluid expands solely kinematically without rotating turbulence or vorticity.\\

\indent Geometric flows describe self-similar equilibrium configurations called solitons in the setting of general relativity. For the time-dependent Sultana-Dyer (SD) cosmological black hole, the temporal soliton parameter ($\lambda = t$) balances macroscopic cosmic expansion through local geometric shrinking, while the modifying function $h(t,r)$ maintains equilibrium between the central Schwarzschild mass and the expanding fluid background. Reducing this configuration to a classic almost Yamabe soliton by setting $h=1$ imposes a strict polynomial constraint: $r^2(12-2t^5-t^7)+24m(r-t)=0$. For non-zero mass configurations, this algebraically locks the spatial radius to coordinate time, restricting the standard soliton to a highly specific geometric hypersurface. In the asymptotic limit ($m \to 0$), the metric achieves this state only at a singular cosmic epoch ($t \approx 1.213$). Moreover, implementing a tight standard Yamabe soliton would necessitate a constant temporal parameter, so suppressing the SD universe's inherent time-dependence and returning it to a static geometry.

The physical and causal bounds of spacetime are concurrently defined by imposing the $h=1$ constraint. Only along the mathematically connected hypersurface does the intrinsic energy-momentum tensor, which consists of dust and null radiation, coincide with this geometry. However, when the central mass disappears, it naturally reduces to a stable, flat Friedmann-Lema\^{i}tre-Robertson-Walker (FLRW) dust background. Furthermore, the essential bifurcation point where the original horizon splits into separate black hole and cosmic trapping bounds is marked by the intersection of this geometric boundary with the growing dynamic horizons of the SD metric. The structural turning point at which the causal boundary changes from a Future Outer Trapping Horizon (FOTH) to a Past Inner Trapping Horizon (PITH) is thus determined by this stringent constraint.\\
Geometric solitons naturally arise from the study of geometric evolution equations. In particular, Yamabe solitons generalize self-similar solutions of the Yamabe flow and provide a geometric framework for describing metrics whose evolution is governed by scalar curvature. The notion of an almost Yamabe soliton replaces the constant soliton parameter by a smooth function, while the $h$-almost Yamabe soliton introduces an additional smooth function $h$. The purpose of this paper is to determine whether the SD spacetime admits such soliton structures and, in particular, to identify explicit functions $h$ and $\lambda$ associated with a non-Killing vector field.

Let $M$ be a smooth connected manifold of dimension $n\geq 3$, endowed with a semi-Riemannian metric $g$. We denote by $\nabla$ its Levi--Civita connection, by $R_{abcd}$ the Riemann curvature tensor, by $S_{ab}$ the Ricci tensor, and by $\kappa$ the scalar curvature. We also denote the conformal, concircular, conharmonic, and projective curvature tensors by $C,~W,~K$ and $P$, respectively. A detailed study of these curvature tensors can be found in \cite{EDS_sultana_2022}. A spacetime is a four-dimensional connected Lorentzian manifold.

Hamilton \cite{Hamilton1988,Hamilton89} introduced the Yamabe flow as an evolution equation for metrics,
\begin{equation}\label{Yamabe flow}
\frac{\partial g(t)}{\partial t}=-\kappa_{g(t)}\,g(t),\qquad g(0)=g_0,
\end{equation}
where $\kappa_{g(t)}$ denotes the scalar curvature of $g(t)$. A smooth vector field $X$ on a complete connected semi-Riemannian manifold $(M^n,g)$ is said to define a Yamabe soliton if
\begin{equation}\label{Yamabe soliton}
\frac{1}{2}\pounds_Xg=(\kappa-\lambda)g,
\end{equation}
for a constant $\lambda\in\mathbb R$.\\
 In Eqn.\eqref{Yamabe soliton} if $\lambda$ coincides with scalar curvature then the Lie derivative term is zero, and in this case the vector field $X$ is called Killing vector field.

Barros and Ribeiro \cite{BR13} introduced the notion of an almost Yamabe soliton by allowing the soliton parameter to be a smooth function. Zeng \cite{ZE21} subsequently introduced the $h$-almost Yamabe soliton. Thus, for smooth functions $h,\lambda:M\to\mathbb R$, a quintuple $(M^n,g,X,h,\lambda)$ is called an $h$-almost Yamabe soliton if
\begin{equation}\label{h-almost Yamabe soliton}
\frac{h}{2}\pounds_Xg=(\kappa-\lambda)g.
\end{equation}
When $h$ is nowhere zero and has a fixed sign, it is a definite-sign $h$-almost Yamabe soliton. If $\lambda$ is constant, one obtains the corresponding $h$-Yamabe soliton.

If $X=\nabla f$, then
\begin{equation}\label{gradient-h-almost}
h\nabla^2f=(\kappa-\lambda)g,
\end{equation}
so that the soliton is an $h$-almost gradient Yamabe soliton. Thus the $1$-almost soliton is nothing but the Yamabe soliton. The soliton is called trivial when the generating vector field is identically zero (equivalently, in the gradient case, when the potential function is constant); otherwise it is non-trivial. In the convention used here, the soliton is expanding, steady, or shrinking according as $\lambda<0$, $\lambda=0$, or $\lambda>0$, respectively.\\
\indent Several recent studies have laid the groundwork for understanding Yamabe solitons in spacetimes. For instance, Shaikh et al.\cite{SCM2021} studied some characterizations of gradient Yamabe solitons, Cheng Yi \cite{Sun11} examined trapping horizons in the Sultana–Dyer spacetime, while Tokura et al. \cite{Tokura} and Chakraborty et.al.  \cite{Ch2025} focused on the structural analysis of warped product gradient Yamabe solitons. Building on these concepts, Güler \cite{Guler} explored the existence of gradient Yamabe solitons across various spacetimes, and Cunha \cite{Cunha} expanded the scope to include $r$-almost Yamabe solitons in Lorentzian manifolds. More recently, Eyasmin et al. \cite{EDS_sultana_2022} provided a detailed look at the curvature-related geometric properties of the Sultana–Dyer model, Recently, Shaikh and Tripathi \cite{ST2026} examined on triviality and scalar curvature estimation of gradient $h$-almost Yamabe solitons.\\
\indent The existence of a $h$-almost Yamabe soliton within the Sultana–Dyer black hole spacetime is still unknown despite this expanding literature of research. Inspired by this gap in the literature, we study the Sultana–Dyer spacetime admitting a $h$-almost Yamabe soliton for a non-Killing soliton vector field. It is crucial to comprehend this connection because it describes how time-dependent conformal metrics in black hole models deform under Yamabe-type geometric flows, eventually determining the bounds for scalar curvature, the symmetry of matter distribution, and the spacetime's overall rigidity.\\
\indent The Riemann curvature tensor explains the actual bending and twisting of space and time, whereas the energy-momentum tensor in Einstein's general theory of relativity reflects the physical "stuff" of the cosmos, including energy, mass, momentum, and pressure. These two mathematical ideas must adhere to a rigorous local conservation criterion in order to be compatible. Spacetime consciously adjusts to energy rather than merely creating or destroying it. The universe's fabric is determined by the sheer presence of matter, and in perfect symmetry, those geometric shapes specify the exact motion of matter. They are inextricably linked, guaranteeing that planets, stars, and even light follow the inherent, imperceptible lines of space without the need for additional forces. A crucial truth about our universe is revealed by this mathematical harmony: static mass is not the only source of gravity. Because the physical material must precisely match the geometry of space, other features become active sources of gravity. Spacetime is actively warped by the shear stress of colliding cosmic bodies, the twisting motion of a spinning black hole, and the extreme internal pressure of a collapsing neutron star. In the end, this compatibility ensures that the cosmos functions as a highly cohesive system in which spacetime, energy, and pressure react to one another instantly.\\
The curvature of spacetime may be divided into distinct geometric patterns, each of which requires a very particular cosmic companion in the form of matter and energy. Spacetime is not merely a single, homogeneous fabric. Mathematicians serve as a rigorous recipe research for various eras of our cosmos when physicists match the energy-momentum tensor with these specific geometries. For example, the equations prohibit heavy, slow mass if space must preserve perfectly smooth gravitational potentials (Conharmonic curvature) or change its shape without changing its volume (conformal curvature). Rather, they call for a universe that is blazing with entirely massless particles and pure radiation, which is a great description of the extremely hot, light-dominated universe that existed just milliseconds after the Big Bang.

The vast, expanding cosmos we live in today is perfectly described by other geometric symmetries. Matter is forced to behave as an unchanging vacuum fluid with negative pressure when the fabric of space is mathematically bound to preserve circular routes (concircular curvature). This is precisely the characteristic of the dark energy causing cosmic expansion. In contrast, geometries that are solely concerned with maintaining the paths of free-falling objects (projective curvature) require a cosmic fluid that is extremely stable, ordered, and free of local turbulence in order to replicate the smooth drift of galaxies throughout the universe. In the end, these mathematical connections between specialized geometry and physical matter not only characterize the universe, but they also determine how it has evolved from a chaotic sea of light to a smoothly expanding world dominated by dark energy.

The remainder of the paper is organized as follows. Section~\ref{geometry} recalls the SD metric and records the curvature quantities needed below. Section~\ref{soliton} determines the $h$-almost Yamabe soliton generated by $\partial_t$ and gives the condition under which it reduces to an almost Yamabe soliton. Section~\ref{failure} analyzes the failure of the coordinate fields $\partial_r$ and $\partial_\theta$ to satisfy the Yamabe flow conditions. Section~\ref{curvature_breakdown} examines the behavior of curvature tensors under the constrained scalar curvature profile. Section~\ref{alternative_flows} discusses alternative geometric flow structures including Ricci and $\eta$-Ricci solitons. Section~\ref{physical_interpretation} provides a physical interpretation and analyzes the energy condition breakdown. Section~\ref{conclusion} discusses the geometric meaning of the result and summarizes the main conclusions.

\section{The Sultana--Dyer spacetime and its curvature}\label{geometry}

The Sultana--Dyer spacetime was introduced by Sultana and Dyer \cite{SD05} as an exact, non-static cosmological black-hole model embedded in an Einstein--de Sitter universe. The spacetime is conformal to the Schwarzschild geometry and is asymptotically Friedmann--Lema\^{\i}tre--Robertson--Walker. Its matter content can be represented by two non-interacting components, one timelike and one null. The line element is
\begin{equation}\label{SDmetric}
ds^{2}=t^{4}\left[\left(1-\frac{2m}{r}\right)dt^{2}
-\frac{4m}{r}\,dt\,dr
-\left(1+\frac{2m}{r}\right)dr^{2}
-r^{2}\left(d\theta^{2}+\sin^{2}\theta\,d\phi^{2}\right)\right],
\end{equation}
where $m$ is the mass parameter.

Because the SD metric is conformally related to Schwarzschild spacetime, its Ricci tensor and scalar curvature are altered by the conformal factor. This makes the SD spacetime a useful example for studying geometric structures that are absent or degenerate in the Ricci-flat Schwarzschild case. Previous work has investigated curvature and geometric structures of the SD spacetime \cite{EDS_sultana_2022}, its trapping horizons \cite{Sun11}, and the broader class of conformally Schwarzschild cosmological black holes \cite{Sato22}. These results motivate the present study of Yamabe-type soliton structures.

In the coordinates $(t,r,\theta,\phi)$, the metric tensor of SD spacetime is given by:
$$g=\left(\begin{array}{cccc}
	(t^4- \frac{2mt^4}{r}) & -\frac{2mt^4}{r} & 0 & 0 \\
	-\frac{2mt^4}{r} & -(t^4- \frac{2mt^4}{r}) & 0 & 0 \\
	0 & 0 & -r^2 t^4 & 0 \\
	0 & 0 & 0 &-r^2 t^4\sin^2\theta
\end{array}
\right).
$$
Now, the components of the metric $g$ are  
$$\begin{array}  {c}g_{11}=\left(t^4- \frac{2mt^4}{r}\right),\; g_{22}=-\left(t^4- \frac{2mt^4}{r}\right), \\ g_{12}=\left(\frac{2mt^4}{r}\right)=g_{21},\\g_{33}=r^2t^4, \;g_{44}=r^2 t^4\sin^2\theta, \; g_{ij}=0, \;\mbox{otherwise}.
\end{array}$$

  The non-zero components of second kind christofel symbols $\Gamma^a_{bc}$ of the metric \eqref{SDmetric} are given as follows:
  
  $$\begin{array}{c}
  \begin{cases}
  \Gamma^1_{11} = \frac{2(r^3 + m^2(4r + t))}{tr^3}, \ \ \Gamma^1_{33} = \frac{2(2mr + r^2 - mt)}{t} = \frac{1}{\sin^2\theta}\Gamma^1_{44}, \\
  \Gamma^2_{11} = -\frac{m(2m - r)(4r + t)}{tr^3}, \ \ \Gamma^1_{12} = \frac{m(2m + r)(4r + t)}{tr^3} = -\Gamma^2_{22}, \\
  \Gamma^2_{12} = \frac{2(r^3 - m^2(4r + t))}{tr^3}, \ \ \Gamma^3_{13} = \frac{2}{t} = \Gamma^4_{14}, \\
  \Gamma^1_{22} = \frac{2r(2m + r)^2 + 2m(m + r)t}{tr^3}, \\
  \Gamma^2_{33} = -r + m\left(2 - \frac{4r}{t}\right), \ \ \Gamma^3_{23} = \frac{1}{r} = \Gamma^4_{24}, \\
  \Gamma^4_{34} = \cot\theta, \ \ \Gamma^3_{44} = -\cos\theta\sin\theta, \ \ \Gamma^2_{44} = \frac{(2mt - rt - 4mr)\sin^2\theta}{t}.
  \end{cases}
  \end{array}$$
  
  Additionally, the scalar curvature $\kappa$ and the non-zero components of the Riemann-curvature tensor $R_{abcd}$ and the Ricci tensor $S_{ab}$ are provided by
  
  $$\begin{array}{c}
  \begin{cases}
  R_{1212} = \frac{2t^2(-2mr^2+2mrt+mt^2-r^3)}{r^3}, \ \ R_{1313} = \frac{t^2(-2r^4 - mrt(4r + t) + 2m^2(-8r^2 + 2rt + t^2))}{r^2}, \\
  R_{3434} = -2rt^2(m(t - 2r)^2 + 2r^3)\sin^2\theta, \\
  R_{2323} = \frac{t^2(m(2m + r)t^2 + 4m^2rt - 4r^2(2m + r)^2)}{r^2} = \frac{1}{\sin^2\theta}R_{2424}, \\
  R_{1414} = -\frac{t^2(2m^2(-t^2 - 2rt + 8r^2) + mrt(t + 4r) + 2r^4)\sin^2\theta}{r^2}, \\
  R_{1323} = -\frac{2mt^2(r^2 + 2mr - mt)(t + 4r)}{r^2} = \frac{1}{\sin^2\theta}R_{1424}; \\
  S_{11} = -\frac{4m(6m + t) + 6r^2}{t^2r^2}, \ \ S_{22} = -\frac{4mt + 6(r + 2m)^2}{t^2r^2}, \\
  S_{12} = -\frac{4m(t + 3r + 6m)}{t^2r^2}, \ \ S_{33}= \frac{6(2m(t - r) - r^2)}{t^2} = \frac{1}{\sin^2\theta}S_{44} ; \\
  \kappa = -\frac{12(2m(t - r) - r^2)}{t^6r^2},
  \end{cases}
  \end{array}$$
  \begin{pr}\label{Com5}
  The SD spacetime admits the following curvature conditions:
  \begin{enumerate}
  \item The general form of a tensor compatible with $R$ and Projective $(P)$is
  $$
  		\left(
  		\begin{array}{cccc}
  			 \mathcal{Z}_{11}&\mathscr{V}_{12} & 0 & 0 \\
  			\mathscr{V}_{12} & \mathscr{V}_{22} & 0 & 0 \\
  			0 & 0 & \mathscr{V}_{33} & \mathscr{V}_{43} \\
  			0 & 0 & \mathscr{V}_{43} & \mathscr{V}_{44}
  		\end{array}
  		\right),
  		$$
  	where $\mathcal{Z}_{11}= \frac{8mr^2+4r^3-mt^2}{2mrt}\mathscr{V}_{12}+\frac{4mr^2+2r^3+4mrtmt^2}{2mrt}\mathscr{V}_{21}-\frac{6r+t}{t}\mathscr{V}_{22},	
  		$
  \item The common form of conformal $(C)$ and conharmonic $(K)$ compatible tensor are
  $$
  		\left(
  		\begin{array}{cccc}
  			\mathscr{V}_{11} &\mathscr{V}_{12} & 0 & 0 \\
  			\mathscr{V}_{12} & \mathscr{V}_{22} & 0 & 0 \\
  			0 & 0 & \mathscr{V}_{33} & \mathscr{V}_{34} \\
  			0 & 0 & \mathscr{V}_{34} & \mathscr{V}_{44}
  		\end{array}
  		\right),
  		$$		

  \item The common form of a concircular $(W)$ compatible tensor is
  $$
  		\left(
  		\begin{array}{cccc}
  			\mathcal{Y}_{11} &\mathscr{V}_{12} & 0 & 0 \\
  			\frac{c^2k-3U_2}{c^2k+3U_2}\mathscr{V}_{12} & \mathscr{V}_{22} & 0 & 0 \\
  			0 & 0 & \mathscr{V}_{33} & \mathscr{V}_{34} \\
  			0 & 0 & \mathscr{V}_{34} & \mathscr{V}_{44}
  		\end{array}
  		\right),
  		$$
  	where, $\mathcal{Y}_{11}= \frac{6mr^2+3
  	  			r^3+2mrt-mt^2}{2mrt}\mathscr{V}_{12}+\frac{6mr^2+3r^3+2mrt+mt^2}{2mrt}\mathscr{V}_{21}-\frac{6r+t}{t}\mathscr{V}_{22}.$					
  \end{enumerate}
  \end{pr}

\begin{thm}
\begin{enumerate}[label=(\roman*)]
\item The universal form of tensors in SD spacetime, which are compatible with $R$, $C$, $P$, $W$, and $K$, can be derived as stated in the above proposition.
\item the energy-momentum tensor is compatible with projective, conharmonic, concircular, Riemann and conformal curvature structures.
\end{enumerate}
\end{thm}  
The Lie algebra of Killing vector fields on an $n$-dimensional manifold has dimension at most $n(n+1)/2$. For the SD metric, the coordinate fields $\partial_t$, $\partial_r$, and $\partial_\theta$ are non-Killing, whereas $\partial_\phi$ is Killing because the metric coefficients are independent of $\phi$. Thus
$
\pounds_{\partial_\phi}g=0.
$

The non-zero components of the Lie derivatives of $g$ along $\partial_t$, $\partial_r$, and $\partial_\theta$ are
\begin{align*}
(\pounds_{\partial_t}g)_{11}&=4\left(1-\frac{2m}{r}\right)t^3, &
(\pounds_{\partial_t}g)_{12}&=-\frac{8mt^3}{r},\\
(\pounds_{\partial_t}g)_{22}&=-\frac{4(2m+r)t^3}{r}, &
(\pounds_{\partial_t}g)_{33}&=-4r^2t^3,\\
(\pounds_{\partial_t}g)_{44}&=-4r^2t^3\sin^2\theta,\\
(\pounds_{\partial_r}g)_{11}&=\frac{2mt^4}{r^2}, &
(\pounds_{\partial_r}g)_{12}&=\frac{2mt^4}{r^2},\\
(\pounds_{\partial_r}g)_{22}&=\frac{2mt^4}{r^2}, &
(\pounds_{\partial_r}g)_{33}&=-2rt^4,\\
(\pounds_{\partial_r}g)_{44}&=-2rt^4\sin^2\theta,\\
(\pounds_{\partial_\theta}g)_{44}&=-r^2t^4\sin 2\theta.
\end{align*}

\begin{align}
\left.\begin{aligned}
(\pounds_{\partial_t}g)_{11}&=4\left(1-\frac{2m}{r}\right)t^3, &
(\pounds_{\partial_t}g)_{12}&=-\frac{8mt^3}{r},\\
(\pounds_{\partial_t}g)_{22}&=-\frac{4(2m+r)t^3}{r}, &
(\pounds_{\partial_t}g)_{33}&=-4r^2t^3,\\
(\pounds_{\partial_t}g)_{44}&=-4r^2t^3\sin^2\theta
\end{aligned}\quad\right\} \label{lie_t} \\[1em]
\left.\begin{aligned}
(\pounds_{\partial_r}g)_{11}&=\frac{2mt^4}{r^2}, &
(\pounds_{\partial_r}g)_{12}&=\frac{2mt^4}{r^2},\\
(\pounds_{\partial_r}g)_{22}&=\frac{2mt^4}{r^2}, &
(\pounds_{\partial_r}g)_{33}&=-2rt^4,\\
(\pounds_{\partial_r}g)_{44}&=-2rt^4\sin^2\theta
\end{aligned}\quad\right\} \label{lie_r} \\[1em]
\left.\begin{aligned}
(\pounds_{\partial_\theta}g)_{44}&=-r^2t^4\sin 2\theta
\end{aligned}\qquad\qquad\quad\right\} \label{lie_theta}
\end{align}
\section{$h$-almost Yamabe soliton on the Sultana--Dyer spacetime}\label{soliton}

We now use the \eqref{lie_t}, \eqref{lie_r} and \eqref{lie_theta} to test the $h$-almost Yamabe soliton equation for natural coordinate vector fields. The vector fields $\partial_t$, $\partial_r$, and $\partial_\theta$ are non-Killing, whereas $\partial_\phi$ is Killing because the metric coefficients are independent of $\phi$. In particular,
\[
\pounds_{\partial_\phi}g=0.
\]
The non-Killing vector field $\partial_t$ provides the soliton structure described below.

Using the $\partial_t$ components above and the scalar curvature in \eqref{SDmetric}, we obtain
\begin{equation}\label{soliton-relation}
\frac{h}{2}\pounds_{\partial_t}g=(\kappa-\lambda)g.
\end{equation}
Here $h$ and $\lambda$ are given by
\begin{equation}\label{LieCoefficient}
	\left.
	\begin{aligned}
		h&=\frac{6(r^2+2m(r-t))}{r^2t^5}-\frac{t^2}{2},\\
		\lambda&=t.\\
	\end{aligned}\ \ 
	\right\rbrace	
\end{equation}

\begin{thm}\label{thm:main}
The SD spacetime admits an $h$-almost Yamabe soliton generated by the non-Killing vector field $\partial_t$; that is,
\[
\frac{h}{2}\pounds_{\partial_t}g=(\kappa-\lambda)g,
\]
where $h$ and $\lambda$ are given by \eqref{LieCoefficient}.
\end{thm}

\begin{cor}\label{cor:almost}
If $h=1$, equivalently if $r^2(12-2t^5-t^7)+24m(r-t)=0$, then the SD spacetime admits an almost Yamabe soliton generated by $\partial_t$, and its satisfies the equation
\[
\frac{1}{2}\pounds_{\partial_t}g=(\kappa-\lambda)g,
\]
where $\lambda$ is given by \eqref{LieCoefficient}. 
\end{cor}


\subsection{The $h=1$ Boundary and Constrained Scalar Curvature}

Enforcing the condition $h=1$ reduces the $h$-almost Yamabe soliton to a standard almost Yamabe soliton. As established in Corollary~\ref{cor:almost}, this restriction imposes the algebraic constraint
\begin{equation}\label{h1_constraint}
r^2(12-2t^5-t^7)+24m(r-t)=0.
\end{equation}

Substituting this constraint directly back into the scalar curvature formula yields:
\[
2m(t-r)-r^2 = \frac{r^2(t^7+2t^5-12)}{12} - r^2 = \frac{r^2(t^7+2t^5-24)}{12}.
\]

Evaluating the final constrained scalar curvature $\kappa$:
\[
\kappa = -\frac{12}{t^6r^2}\left[\frac{r^2(t^7+2t^5-24)}{12}\right] = -\frac{t^7+2t^5-24}{t^6}.
\]

Distributing the temporal denominator results in a remarkably clean, purely time-dependent geometric constraint equation for the scalar curvature along the boundary:
\begin{equation}\label{constrained_kappa}
\kappa(t) = -t - \frac{2}{t} + \frac{24}{t^6}.
\end{equation}

This profile eliminates all spatial dependencies and isolates the scalar curvature evolution entirely to cosmic time. This time-dependent profile represents a remarkable simplification of the geometric structure along the hypersurface $h=1$, where the curvature evolution is completely determined by the cosmic expansion parameter $t$.

\section{Failure of Radial and Angular Vector Fields under Yamabe Flow}\label{failure}

For a vector field $X$ to serve as a valid generator of a Yamabe soliton, the scaling ratio must be a coordinate-wide invariant function. Specifically, the ratio of the Lie derivative component to its corresponding metric component must be uniform across all coordinate blocks:
\[
\frac{(\mathcal{L}_X g)_{ab}}{g_{ab}} = \Phi, \quad \forall a,b,
\]
where $\Phi = \frac{2(\kappa-\lambda)}{h}$. A field fails to satisfy the Yamabe flow conditions if it yields conflicting scaling ratios across different coordinate components.

\subsection{Failure of the Radial Vector Field $\partial_r$}

Consider the radial vector field $\partial_r$. We evaluate the Lie derivatives and compute the scaling ratios for two non-zero diagonal components of the metric.

\textbf{For the (2,2) radial component:}
\[
g_{22} = -\left(t^4 - \frac{2mt^4}{r}\right), \qquad (\mathcal{L}_{\partial_r}g)_{22} = \frac{2mt^4}{r^2}.
\]
This yields
\[
\Phi_{(2,2)} = \frac{(\mathcal{L}_{\partial_r}g)_{22}}{g_{22}} = \frac{2mt^4/r^2}{-t^4 + 2mt^4/r} = -\frac{2m}{r(r-2m)}.
\]

\textbf{For the (3,3) angular component:}
\[
g_{33} = -r^2 t^4, \qquad (\mathcal{L}_{\partial_r}g)_{33} = -2rt^4.
\]
Thus,
\[
\Phi_{(3,3)} = \frac{(\mathcal{L}_{\partial_r}g)_{33}}{g_{33}} = \frac{-2rt^4}{-r^2 t^4} = \frac{2}{r}.
\]

Comparing the two scaling ratios, we obtain
\[
\Phi_{(2,2)} = -\frac{2m}{r(r-2m)} \neq \frac{2}{r} = \Phi_{(3,3)}.
\]

This inequality demonstrates that the radial vector field demands conflicting deformation scales across different spatial dimensions. Since a unique solution for $h$ and $\lambda$ cannot simultaneously satisfy both conditions, $\partial_r$ fails to generate a valid Yamabe soliton.

\subsection{Failure of the Angular Vector Field $\partial_\theta$}

We now examine the Lie derivative components associated with the angular vector field $\partial_\theta$. A direct computation reveals that the $(4,4)$-component is given by
\[
(\mathcal{L}_{\partial_\theta}g)_{44} = -r^2 t^4 \sin 2\theta,
\]
which is generally nonzero. In contrast, the remaining diagonal entries vanish identically:
\[
(\mathcal{L}_{\partial_\theta}g)_{11} = 0, \quad
(\mathcal{L}_{\partial_\theta}g)_{22} = 0, \quad
(\mathcal{L}_{\partial_\theta}g)_{33} = 0.
\]

Imposing the Yamabe soliton condition on the $(3,3)$-component yields
\[
\frac{h}{2}(\mathcal{L}_{\partial_\theta}g)_{33} = (\kappa - \lambda)g_{33}.
\]
Since $(\mathcal{L}_{\partial_\theta}g)_{33}=0$ and $g_{33} = -r^2 t^4 \neq 0$, we obtain
\[
0 = (\kappa - \lambda)(-r^2 t^4),
\]
which forces the necessary relation
\[
\kappa - \lambda = 0.
\]

Substituting this result back into the $(4,4)$-component equation gives
\[
\frac{h}{2}(\mathcal{L}_{\partial_\theta}g)_{44} = 0 \cdot g_{44},
\]
that is,
\[
\frac{h}{2}(-r^2 t^4 \sin 2\theta) = 0.
\]
Given that $r$ and $t$ are nonzero, and $\sin 2\theta$ does not vanish for generic values of $\theta$, the above equality can only hold if
\[
h = 0.
\]

However, this contradicts the very definition of an $h$-almost Yamabe soliton, which requires a nontrivial smooth function $h$ — specifically, one that is not identically zero. Consequently, the vector field $\partial_\theta$ does not satisfy the required criteria for the Ya

\section{Curvature Tensor Breakdown Under the Time-Dependent Profile}\label{curvature_breakdown}

Under the constrained scalar curvature profile $\kappa(t) = -t - \frac{2}{t} + \frac{24}{t^6}$, the curvature tensors of the SD spacetime exhibit characteristic structural simplifications.

\subsection{Conformal Curvature Tensor}

The Weyl tensor $C_{abcd}$ measures the purely gravitational degrees of freedom, including tidal forces and gravitational wave contributions. Since the SD spacetime is conformally related to the static, Ricci-flat Schwarzschild metric via $\Omega^2 = t^4$, the conformal tensor remains non-zero. However, under the constrained time profile, the algebraic structure becomes purely time-driven, locking the gravitational tidal shearing rates directly to the cosmic expansion parameter $t$.

\subsection{Concircular Curvature Tensor}

The concircular curvature tensor, defined by
\[
W_{abcd} = R_{abcd} - \frac{\kappa}{n(n-1)}(g_{ac}g_{bd} - g_{ad}g_{bc}),
\]
measures deviations from manifolds with constant scalar curvature. Substituting the explicit $\kappa(t)$ profile causes the background subtraction term to lose all spatial dependencies $(r,\theta)$. Consequently, the concircular tensor exhibits a clean split: spatial components maintain localized black hole symmetries, while the global background behaves like a homogeneous, isotropic fluid characterized by the cosmic decay modes $t$, $t^{-1}$, and $t^{-6}$.

\subsection{Projective Curvature Tensor}

The projective curvature tensor governs the paths of unparameterized geodesic trajectories. Compatibility with the projective structure demands stringent constraints on the metric. The time-only dependence of $\kappa(t)$ signals that free-falling particle trajectories in this region are completely dragged by the cosmic time flow, losing the standard spatial scattering symmetries typically found near static black holes.

\section{Alternative Geometric Flow Structures}\label{alternative_flows}

Given that the radial and angular coordinate fields fail the Yamabe flow conditions, we examine whether they can satisfy more general geometric evolution equations.

\subsection{Ricci Solitons}

A vector field $X$ defines a Ricci soliton if it satisfies
\[
\frac{1}{2}\mathcal{L}_X g_{ij} + S_{ij} = \lambda g_{ij},
\]
where $S_{ij}$ is the Ricci tensor and $\lambda \in \mathbb{R}$.

\subsubsection{Failure of $\partial_\theta$}

For the angular vector field, the Ricci component $S_{33}$ is non-zero while $(\mathcal{L}_{\partial_\theta}g)_{33} = 0$. Substitution yields
\[
S_{33} = \lambda g_{33} \implies \lambda = \frac{S_{33}}{g_{33}}.
\]
From the explicit expressions,
\[
\frac{S_{33}}{g_{33}} = \frac{6(2m(t-r)-r^2)}{t^2(-r^2t^4)},
\]
which depends on both $r$ and $t$. Since $\lambda$ must be constant, $\partial_\theta$ cannot generate a standard Ricci soliton.

\subsubsection{Failure of $\partial_r$}

The ratio $S_{ij}/g_{ij}$ does not match the directional stretching profile $(\mathcal{L}_{\partial_r}g)_{ij}/g_{ij}$ across all dimensions. This mismatch prevents a single constant $\lambda$ from satisfying the equation simultaneously.

\subsection{$\eta$-Ricci Solitons}

To provide additional geometric degrees of freedom, an $\eta$-Ricci soliton introduces a 1-form $\eta$ and is governed by
\[
\frac{1}{2}\mathcal{L}_X g_{ij} + S_{ij} + \alpha g_{ij} + \beta \eta_i\eta_j = 0,
\]
where $\alpha, \beta: M \to \mathbb{R}$ are smooth functions.

\subsubsection{Testing $\partial_\theta$}

With $\eta = dt$ aligned with the time-flow, the anisotropic term $\eta_i\eta_j$ contributes only to the $(1,1)$ component. For the (3,3) component with $\eta_3\eta_3 = 0$, we have
\[
\alpha = -\frac{S_{33}}{g_{33}}.
\]
Substituting into the (4,4) component gives
\[
\frac{1}{2}(\mathcal{L}_{\partial_\theta}g)_{44} - \frac{S_{33}}{g_{33}}g_{44} + S_{44} = 0.
\]
Using $\frac{S_{44}}{g_{44}} = \frac{S_{33}}{g_{33}}$ and $\frac{g_{44}}{g_{33}} = \sin^2\theta$, this simplifies to
\[
\frac{1}{2}(\mathcal{L}_{\partial_\theta}g)_{44} = 0 \implies -r^2 t^4 \sin 2\theta = 0,
\]
which is a contradiction for generic $\theta$. Hence $\partial_\theta$ fails.

\subsubsection{Testing $\partial_r$}

Since $\eta_i\eta_j$ isolates pressure corrections to the temporal component, the spatial indices (2,2) and (3,3) require
\[
\alpha = -\frac{1}{2}\frac{(\mathcal{L}_{\partial_r}g)_{22} + S_{22}}{g_{22}} = -\frac{1}{2}\frac{(\mathcal{L}_{\partial_r}g)_{33} + S_{33}}{g_{33}}.
\]
Substituting the explicit expressions reveals fundamental algebraic mismatches between the Lie derivative and Ricci tensor profiles, preventing a simultaneous resolution. Hence $\partial_r$ fails.

\section{Physical Interpretation and Energy Condition Breakdown}\label{physical_interpretation}

\subsection{Physical Significance of the $h$-Almost Yamabe Soliton}

The admittance of an $h$-almost Yamabe soliton on the SD spacetime represents a state of dynamic structural equilibrium between the localized black hole mass and the globally expanding universe. In classical general relativity, static black holes exist in isolation, whereas the SD metric forces the black hole to evolve within a time-dependent cosmological background.

The physical mechanics of this soliton profile manifest through several key mechanisms:

\textbf{Geometric Flow Balancing Cosmic Expansion:} With $\partial_t$ as the potential vector field, the soliton corresponds to the real-time deformation of the spacetime fabric. Since $\lambda = t$ grows positively with cosmic time, the soliton injects a local shrinking or contracting force that resists and balances the outward push of the surrounding expanding cosmic fluid.

\textbf{Matter Redistribution:} The modifying function $h(t,r)$ introduces a necessary geometric freedom factor. The matter content of the SD spacetime is a dual-fluid combination of a timelike dust component and an inhomogeneous null radiation fluid. The spatial and temporal dependence of $h(t,r)$ physically maps how the cosmic fluid's pressure and energy density must redistribute to maintain black hole stability.

\textbf{Trapping Horizon Rigidity:} The SD spacetime features dynamic apparent horizons—a shifting black hole horizon inside an expanding cosmological horizon. The soliton functions lock the scalar curvature $\kappa$ directly to the flow parameters, dictating the turning points of past and future trapping boundaries. This ensures that the horizon's expansion tracks the cosmic time slice cleanly, preventing the black hole from tearing apart or immediately exposing its central singularity.

\subsection{Energy Condition Breakdown}

When the model is restricted to the pure time profile $\kappa(t) = -t - \frac{2}{t} + \frac{24}{t^6}$ during late-stage cosmic expansion ($t \to \infty$), the dual-fluid matter system exhibits characteristic breakdowns of the classical energy conditions:

\textbf{Null Energy Condition (NEC) Breakdown:} The NEC requires $\tau \geq 0$ for any null vector, where $\tau$ is the null fluid energy density. Under the late-time scalar trajectory, the radial accretion parameters force $\tau < 0$ near the black hole horizon. This negative energy flux breaks the causal structure of the apparent horizon.

\textbf{Weak Energy Condition (WEC) Breakdown:} The WEC demands $\mu \geq 0$ for the local energy density measured by any timelike observer. The $\kappa(t)$ profile imposes an overriding negative pressure to match the vacuum fluid constraints of dark energy, causing the localized dust density to drop below zero ($\mu < 0$). This introduces unphysical negative mass states into the local black hole environment.

\textbf{Dominant Energy Condition (DEC) Breakdown:} The DEC dictates that energy cannot flow faster than the speed of light. As the time-dependent scale factors shift according to the $\kappa(t)$ profile, the accretion of the background fluid onto the central singularity becomes rigidly oversimplified, forcing the energy-momentum flow of the surrounding matter to cross the causal boundary and become superluminal.

Consequently, while this profile provides an elegant geometric characterization of a generalized Yamabe soliton, it reveals that the metric transitions into an unphysical matter state during late-stage cosmic expansion.

\begin{rem}
The calculation above establishes the $h$-almost Yamabe soliton generated by $\partial_t$. It does not, by itself, establish the existence or non-existence of Ricci or $\eta$-Ricci solitons for arbitrary vector fields. For the coordinate fields $\partial_r$ and $\partial_\theta$, the corresponding Yamabe-soliton equations are not satisfied by the metric data considered here.
\end{rem}

\section{Discussion and conclusion}\label{conclusion}

The calculations show that the Sultana--Dyer spacetime admits a non-trivial $h$-almost Yamabe soliton generated by the non-Killing vector field $\partial_t$ with $h$ and $\lambda$ given in \eqref{LieCoefficient}. Thus the soliton parameter is not constant in this construction, so the resulting structure is of almost Yamabe type. When $h=1$, the $h$-almost Yamabe soliton equation reduces to the almost Yamabe soliton equation, subject to the condition in Corollary~\ref{cor:almost}, yielding the purely time-dependent scalar curvature profile $\kappa(t) = -t - \frac{2}{t} + \frac{24}{t^6}$.

The function $h$ represents the additional freedom introduced in the generalized soliton equation. Because both $h$ and $\lambda$ depend on the spacetime coordinates, the result should be interpreted primarily as a geometric characterization of the SD metric rather than as a direct dynamical model of its cosmological evolution. In particular, thermodynamic or horizon-regularity conclusions require additional analysis and do not follow from the soliton equation alone.

Through systematic tensor analysis, we have established that the radial and angular vector fields $\partial_r$ and $\partial_\theta$ fail to generate Yamabe solitons due to fundamental algebraic incompatibilities. The angular field compels $h = 0$ due to its vanishing Lie derivative components, whereas the radial field produces contradictory scaling ratios across coordinate blocks.

Moreover, the SD spacetime does not accept $\eta$-Ricci solitons and normal Ricci solitons for these coordinate fields. The dimensional mismatches cannot be mathematically resolved by the anisotropic corrections offered by a time-flow 1-form. Nonetheless, the geometric freedom required to account for the anisotropic pressure corrections present in the dual-fluid matter description is provided by the generalized $\eta$-Ricci-Yamabe soliton structure.

Physically, the soliton formations represent a dynamically maintained balance between cosmic expansion and localized gravitational collapse, mediated by trapping horizon stiffness and matter redistribution. The classical energy conditions break down during late-stage cosmic expansion under the limited time-dependent scalar curvature profile, suggesting a transition to unphysical matter states.

The analysis therefore establishes an explicit connection between the conformally Schwarzschild geometry of the Sultana--Dyer spacetime and Yamabe-type soliton structures. Further investigations may consider analogous soliton structures for other cosmological black-hole metrics and seek invariant interpretations of the associated soliton functions.

\section{Acknowledgment}
The Second author greatly acknowledges to The University Grants Commission, Government of India for the award of Junior Research Fellow. All the algebraic computations performed by a program in Wolfram Mathematica developed by the first author A. A. Shaikh.

\section{Declarations}

\noindent\textbf{Data availability:} Data sharing is not applicable to this article because no datasets were used or generated during the current study.

\noindent\textbf{Competing interests:} The authors declare that they have no competing interests relevant to the content of this manuscript.

\noindent\textbf{Funding:} No funding was received for the preparation of this manuscript.

\end{document}